\documentclass[noeprint,twocolumn,prx,showpacs,groupedaddress,superscriptaddress,nofootinbib]{revtex4-2} 
\usepackage{graphicx}  
\usepackage[dvipsnames]{xcolor} 
\usepackage{color}     
\usepackage{amsmath}   
\usepackage{amsfonts}  
\usepackage{float}     
\usepackage[pdftex,linkcolor=blue,citecolor=blue,urlcolor=blue,colorlinks]{hyperref} 
\usepackage{outlines} 
\usepackage{lipsum}
\usepackage{bm}    
\usepackage{mathrsfs}

\begin{document}

\title{Local Hilbert space fragmentation and weak thermalization\\in Bose-Hubbard diamond necklaces}

\author{Eloi Nicolau}
\affiliation{%
	Departament de F\'{i}sica, Universitat Aut\`{o}noma de Barcelona, E-08193 Bellaterra, Spain.}%

\author{Anselmo M. Marques}%
\affiliation{Department of Physics and i3N, University of Aveiro, 3810-193 Aveiro, Portugal.}%

\author{Jordi Mompart}%
\affiliation{%
	Departament de F\'{i}sica, Universitat Aut\`{o}noma de Barcelona, E-08193 Bellaterra, Spain.}%

\author{Ver\`{o}nica Ahufinger}%
\affiliation{%
	Departament de F\'{i}sica, Universitat Aut\`{o}noma de Barcelona, E-08193 Bellaterra, Spain.}%

\author{Ricardo G. Dias}%
\affiliation{Department of Physics and i3N, University of Aveiro, 3810-193 Aveiro, Portugal.}%

\begin{abstract}
We study Bose-Hubbard models in a family of diamond necklace lattices with $n$ central sites. The single-particle spectrum of these models presents compact localized states (CLSs) that occupy the up and down sites of each diamond. By performing an appropriate basis rotation, the fragmentation of the many-boson Hilbert space becomes apparent in the adjacency graph of the Hamiltonian, showing disconnected sub-sectors with a wide range of dimensions. The models present a conserved quantity related to the occupation of the single-particle CLSs that uniquely identifies the different sub-sectors of the many-boson Hilbert space. Due to the fragmentation of the Hilbert space, the distribution of entanglement entropies of the system presents a nested-dome structure. We find weak thermalization through sub-sector-restricted entanglement evolution and a wide range of entanglement entropy scalings from area-law to logarithmic growth. Additionally, we observe how the distinguishability between the different domes increases with the number of central sites and we explain the mechanism behind this fact by analyzing the graph structure of the Hamiltonian. 
\end{abstract}
\maketitle

\section{Introduction}\label{SecIntroduction}
The Eigenstate Thermalization Hypothesis (ETH) predicts how an excited state of a many-body closed quantum system should thermalize \cite{Deutsch1991,Srednicki1994,Rigol2008}. Although most systems obey this hypothesis, numerous examples of non-ergodic systems have been found. Perhaps the most prominent example is integrable systems, where the number of conserved quantities equals or exceeds the degrees of freedom of the system, thus exactly determining all the eigenstates \cite{Sutherland2004}. In many-body localized systems \cite{Abanin2019}, the interplay between disorder and interactions gives rise to emergent integrability, which also leads to a strong violation of the ETH. More recently, it was shown that the ETH can also be weakly violated by a vanishing subset of non-thermal eigenstates, dubbed Quantum Many Body Scars (QMBS). They were initially found in one-dimensional Rydberg arrays \cite{Bernien2017} with the underlying PXP model \cite{Turner2018,Turner2018a}, and were also discovered in parallel in the AKLT model \cite{Affleck1987,Moudgalya2018}. Since these initial works, QMBS have been found in several systems where there is either a tower of scarred eigenstates  \cite{Moudgalya2018a,Moudgalya2018,Iadecola2019,Schecter2019,Iadecola2020,Mark2020,Mark2020a,Moudgalya2020a,Shibata2020,Chattopadhyay2020a,ODea2020,Lee2020,Kuno2021,Jeyaretnam2021} or an isolated scar \cite{Shiraishi2017,Ok2019,Surace2020,McClarty2020,Kuno2020,Srivatsa2020a,VanVoorden2021a,Banerjee2021,Chertkov2021}.

A broader phenomenon that also leads to weak thermalization is Hilbert space fragmentation, also known as Hilbert space shattering or Krylov fracture \cite{Moudgalya2021a}. The Hilbert space presents exponentially many dynamically disconnnected sectors that prevent the system from thermalizing completely. Remarkably, this mechanism can lead both to a weak or a strong violation of the ETH. This effect can arise in a wide variety of systems, such as dipole moment or center-of-mass conserving systems \cite{Sala2020,Khemani2020,Taylor2020,Moudgalya2021c,Doggen2021}, the 1D $t$-$J_z$ model \cite{Rakovszky2020}, the $t$-$V$ and $t$-$V_1$-$V_2$ models \cite{DeTomasi2019,Frey2022a}, and models with dipolar interactions \cite{Li2021}. All the above examples exhibit fragmentation of the Hilbert space in the product state basis \cite{Moudgalya2021b}, \textit{i.e.}, classical fragmentation. Quantum fragmentation, which occurs in an entangled basis, has been recently shown to arise in Temperley-Lieb spin chains \cite{Moudgalya2021b} and in quantum East models \cite{Brighi2022}. However, it has yet to be determined if quantum fragmentation leads to different phenomenology than its classical analogue.

The fragmentation in the above examples has recently been referred to as \textit{standard} Hilbert space fragmentation, to distinguish it from \textit{local} Hilbert space fragmentation \cite{Buca2022}, that arises in models with \cite{McClarty2020,Lee2020,Chertkov2021,Hahn2021,Lee2021} or without \cite{Richter2022} frustration and in flat band models \cite{Danieli2020}. While  standard fragmentation is due to the presence of non-local conserved quantities, locally fragmented systems present strictly local conservation laws.

In this work, we report on a family of Bose-Hubbard diamond necklaces \cite{Kempkes2022} that exhibit quantum local Hilbert space fragmentation. Here, the presence of a single-particle flat band composed of compact localized states (CLSs) gives rise to the fragmentation of the Hilbert space when introducing on-site interactions. As a consequence of this fragmentation, one finds a nested distribution of entanglement entropies, sector-restricted thermalization, and a broad range of sub-sectors of the Hamiltonian that range from frozen sub-sectors following area-law to non-integrable sub-sectors with logarithmic scaling.

The article is structured as follows: in Section \ref{SecPhysicalSystem}, we introduce the system and we describe the basis rotation that reveals the fragmentation of the Hilbert space in Sec. \ref{SecBasisRotation}. In Sec. \ref{SecParities}, we analyze the conserved quantity that characterizes the sub-sectors of the Hamiltonian, discuss the adjacency graphs of the fragmented Hamiltonian, and demonstrate that the system is strongly fragmented. The numerical results are discussed in Sec.~\ref{SecNumericalResults}, which include the distribution of entanglement entropies, the entanglement evolution and scaling, the level spacing analysis and a comparison between the different models of the diamond necklace family. Finally, we summarize our conclusions in Sec. \ref{SecConclusions}.

\begin{figure}[t]
	\includegraphics{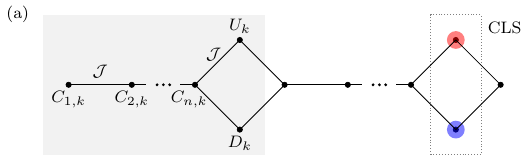}
	\includegraphics{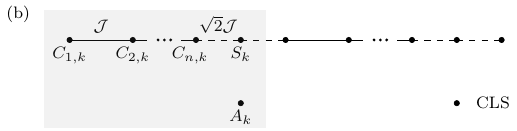}
	\caption{(a) Diagram of the one-dimensional diamond necklace model with $n$ central sites. All couplings have a strength $\mathcal{J}$ and the unit cell is shadowed in gray. In the second unit cell we represent the CLS with the site amplitude being the radius of the circle and the phase being the color (zero, red; $\pi$, blue). (b) Diagram of the rotated model with the renormalized couplings, $\sqrt{2}\mathcal{J}$, denoted by a dashed line. The uncoupled states represent the CLSs, $|A_k\rangle$.}\label{FigPhysicalSystem}
\end{figure}

\section{Physical system}\label{SecPhysicalSystem}
We study a system of interacting bosons loaded onto a one-dimensional lattice of diamond necklaces with $n$ central (\textit{i.e.} spinal) sites [see Fig.~\ref{FigPhysicalSystem}(a)]. Each unit cell $k$ is composed of the sites $C_{1,k}\cdots C_{n,k}$, $U_k$ and $D_k$ (with $k=1,...,N_c$), and all the couplings have the same magnitude $\mathcal{J}$. The Hamiltonian of this system is $\hat{\mathcal{H}}_n=\hat{\mathcal{H}}_n^0+\hat{\mathcal{H}}_n^{int}$, where the single-particle Hamiltonian reads
\begin{equation}\label{EqSingleParticleHamiltonian}
	\begin{aligned}
		\hat{\mathcal{H}}_{n}^0=\mathcal{J} \sum_{k}^{}\Bigg[&\hat{c}^{\dagger}_{n,k}(\hat{u}_{k}+\hat{d}_{k})+(\hat{u}^\dagger_{k}+\hat{d}^\dagger_{k})\hat{c}_{1,k+1}+\\&
		+\sum_{j=1}^{n-1}(\hat{c}^{\dagger}_{j,k} \hat{c}_{j+1,k})\Bigg]+\mathrm{H.C.},
	\end{aligned}
\end{equation}
where $\hat{c}_{j,k}$ is the annihilation operator of the state $|C_{j,k}\rangle$ at the central site $j=1,...,n$ in each unit cell $k$, and $\hat{u}_{k}$ and $\hat{d}_{k}$ are the annihilation operators of the states $|U_{k}\rangle$ and $|D_{k}\rangle$ at the up and down sites of each diamond, respectively. In particular, the $n=2$ case corresponds to a type of orthogonal dimer chain \cite{Ivanov1997a,Richter1998,Koga2000,Honecker2004a,Canova2004,Paulinelli2013,Verkholyak2013,Nandy2019a,Rojas2019,Galisova2021,Zurita2021} with absent vertical couplings. The interaction Hamiltonian reads
\begin{equation}\label{EqInteractionHamiltonianHubbard}
	\begin{aligned}
		\hat{\mathcal{H}}^{int}_{n}=&\dfrac{U}{2}\sum_{k=1}^{N_c}\Bigg[\hat{n}_{u,k}(\hat{n}_{u,k}-1)+\hat{n}_{d,k}(\hat{n}_{d,k}-1)\\
		&+\sum_{j=1}^{n} \hat{n}_{j,k}(\hat{n}_{j,k}-1)\Bigg]=\hat{\mathcal{H}}^{int}_{n,\text{diam.}}+\hat{\mathcal{H}}^{int}_{n,\text{cent.}},
	\end{aligned}
\end{equation}
where we distinguish the terms of the up and down sites of each diamond, $\hat{\mathcal{H}}^{int}_{n,\text{diam.}}$, and the central sites, $\hat{\mathcal{H}}^{int}_{n,\text{cent.}}$.  $\hat{n}_{u,k}=\hat{u}^{\dagger}_{k}\hat{u}_{k}$, $\hat{n}_{d,k}=\hat{d}^{\dagger}_{k}\hat{d}_{k}$ and $\hat{n}_{j,k}=\hat{c}^{\dagger}_{j,k}\hat{c}_{j,k}$ are the number operators at the up, down and central sites, respectively.

An interesting characteristic of this family of Hamiltonians is that each diamond presents a single-particle compact localized state (CLS) that only populates the sites $U_k$ and $D_k$, $\left(|U_{k}\rangle-|D_k\rangle\right)/\sqrt{2}$, [see Fig.~\ref{FigPhysicalSystem}(a)]. Due to the presence of the CLS in each diamond of the lattice, all models of this family exhibit a single-particle spectrum with a zero-energy flat band. We are interested in the many-body states where some of the particles occupy a CLS, and how the existence of these states modifies the thermalization properties of the whole system. The numerical results that we present in Section \ref{SecNumericalResults} can be better interpreted by performing a basis rotation and analyzing the symmetries of the system, which we discuss in the next subsection.

\begin{figure*}[t]
	\includegraphics[width=2.05\columnwidth]{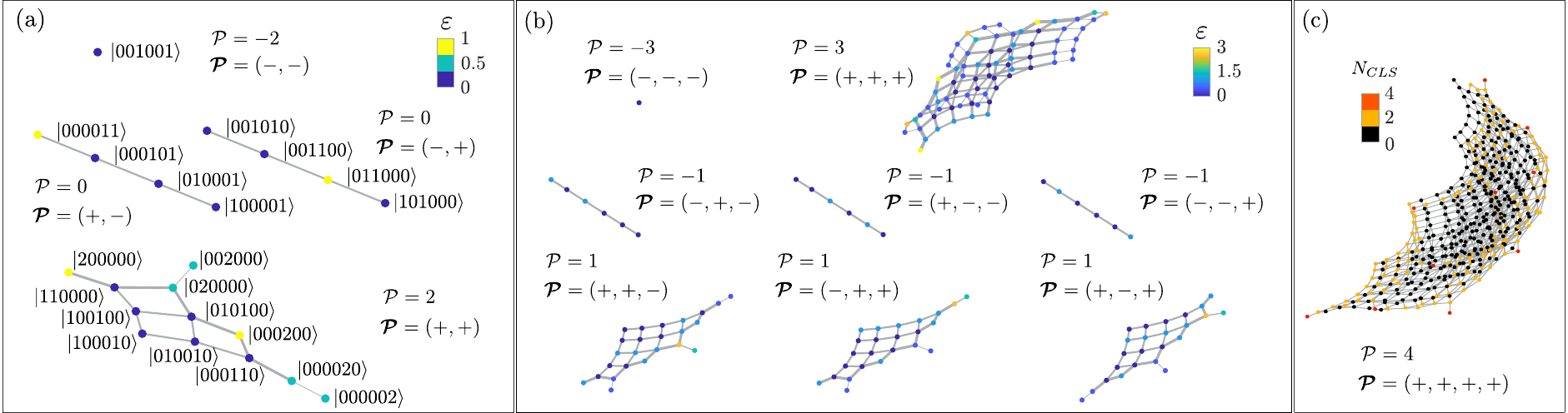}
	\caption{Adjacency graphs for open boundary conditions and $U/\mathcal{J}=1$. (a)  $\hat{\mathcal{H}}_1^\prime$, with $N=2$ particles in $N_c=2$ unit cells, (b) $\hat{\mathcal{H}}_1^\prime$, with $N=3$ and $N_c=3$, (c) largest sub-sector of $\hat{\mathcal{H}}_1^\prime$, with $N=4$ and $N_c=4$. The width of the lines indicates the strength of the couplings between basis states and the color of the nodes represents the diagonal terms, $\varepsilon$, in (a) and (b) and the total number of particles in a CLS,  $N_{CLS}$, in (c). For each cluster, the values of the global CLS number parity are given as well as the vector $\bm{\mathcal{P}}$ for the local CLS number parity. In (a), each basis state is represented by a node and labeled using the notation $|N_{C,1}\,N_{S,1}\,N_{A,1}\,N_{C,2}\,N_{S,2}\,N_{A,2}\rangle$, where $N_{j,k}$ is the number of particles in state $|j_k\rangle$ ($j=C,S,A$) in the unit cell $k$.}\label{FigGraphs}
\end{figure*}

\subsection{Basis rotation}\label{SecBasisRotation}
Consider the symmetric and antisymmetric superpositions of the up and down states of each diamond, 
\begin{equation}\label{EqBasisChange}
	|S_k\rangle=\dfrac{1}{\sqrt{2}}\left(|U_{k}\rangle+|D_k\rangle\right), \quad
	|A_k\rangle=\dfrac{1}{\sqrt{2}}\left(|U_{k}\rangle-|D_k\rangle\right), 
\end{equation}
where $\hat{s}_k^\dagger$ and $\hat{a}_k^\dagger$ are the respective creation operators and $|A_k\rangle$ is the CLS in unit cell $k$. By using these states to perform a basis rotation on the single-particle Hamiltonian, in Eq.~(\ref{EqSingleParticleHamiltonian}), only the couplings associated to the diamonds are altered,
\begin{equation}\label{EqSingleRotatedParticleHamiltonian}
	\begin{aligned}
	\hat{\mathcal{H}}_{n}^{0\prime}= \sum_{k}^{}\Bigg[&\sqrt{2}\mathcal{J}\bigg( \hat{c}^{\dagger}_{n,k}\hat{s}_{k}+\hat{s}^\dagger_{k}\hat{c}_{1,k+1}\bigg)+\\&
	+\mathcal{J}\sum_{j=1}^{n-1}(\hat{c}^{\dagger}_{j,k} \hat{c}_{j+1,k})\Bigg]+\mathrm{H.C.}
	\end{aligned}
\end{equation}
One obtains a linear chain that includes the symmetric states, $|S_k\rangle$, and the central states $|C_{j,k}\rangle$, with renormalized couplings corresponding to the diamonds, $\sqrt{2}\mathcal{J}$. 
Additionally, the CLSs in each unit cell, $|A_k\rangle$, become decoupled, see Fig.~\ref{FigPhysicalSystem}(b). In analogy with the transformation of $\hat{\mathcal{H}}_n^0$, only the interaction term of the up and down sites of each diamond, $\hat{\mathcal{H}}^{int}_{n,\text{diam.}}$ in Eq.~(\ref{EqInteractionHamiltonianHubbard}), is altered by the basis rotation,
\begin{equation}\label{EqRotatedInteractionHamiltonian}
	\begin{aligned}
		\hat{\mathcal{H}}^{int\prime}_{n,\text{diam.}}&=\dfrac{U}{4}\sum_{k=1}^{N_c}\bigg[4\hat{s}_{k}^{\dagger} \hat{a}_{k}^{\dagger} \hat{s}_{k} \hat{a}_{k}+\sum_{\sigma=a,s}\left(\hat{\sigma}_{k}^{\dagger}  \hat{\sigma}_{k}^{\dagger}\hat{\sigma}_{k} \hat{\sigma}_{k}\right) \\
		&+\hat{a}_{k}^{\dagger} \hat{a}_{k}^{\dagger} \hat{s}_{k} \hat{s}_{k}+\hat{s}_{k}^{\dagger}\hat{s}_{k}^{\dagger} \hat{a}_{k} \hat{a}_{k}\bigg],
	\end{aligned}
\end{equation}
where $\hat{\sigma}_{k}$ ($\hat{\sigma}=\hat{s},\hat{a}$) are the annihilation operators of $|S_k\rangle$ and $|A_k\rangle$, respectively. The first term corresponds to a nearest-neighbor interaction that arises when there is at least one particle in $|S_{k}\rangle$ and one in $|A_{k}\rangle$, akin to the inter-circulation interaction term appearing in Hubbard models of excited orbital angular momentum states in optical lattices \cite{Pelegri2020,Nicolau2022}. The second term is an effective on-site interaction that occurs when there are at least two particles in either $|S_{k}\rangle$ or $|A_{k}\rangle$. Finally, the last two terms correspond to a two particle tunnelling between the decoupled states $|A_{k}\rangle$ and the states $|S_{k}\rangle$. Therefore, on-site interactions induce a coupling between the CLSs and the dispersive linear chain through the two-particle tunnelling.

\subsection{Local and global CLS number parity}\label{SecParities}
Let us consider the two-particle tunnelling term that appears in the rotated interaction Hamiltonian of Eq.~(\ref{EqRotatedInteractionHamiltonian}). As a consequence of this process, the system presents a conserved quantity, the local CLS number parity, that reads
\begin{equation}
	\hat{\mathcal{P}}_k=e^{i\pi \hat{n}_{a,k}},
\end{equation}
where $\hat{n}_{a,k}=\hat{a}_k^\dagger \hat{a}_k$ is the CLS number operator at unit cell $k$. This operator commutes with the rotated interaction Hamiltonian, $[\hat{\mathcal{H}}^{int\prime}_{n,\text{diam.}},\hat{\mathcal{P}}_k]=0$, and consequently, with the total rotated Hamiltonian, $[\hat{\mathcal{H}}_n^\prime,\hat{\mathcal{P}}_k]=0$. The operator $\hat{\mathcal{P}}_k$ can be evaluated at each unit cell $k$ (which contains a single diamond) and takes the eigenvalues $\mathcal{P}_k=1$, for an even number of particles, and $\mathcal{P}_k=-1$, for an odd number of particles. We define the local CLS number parity vector as the vector that contains the eigenvalues of $\hat{\mathcal{P}}_k$ at each unit cell, $\bm{\mathcal{P}}=(\mathcal{P}_1,\cdots,\mathcal{P}_{N_c})$. This conserved quantity corresponds to a $\mathbb{Z}_2$ local gauge  symmetry governed by the two-particle tunnelling term in Eq.~(\ref{EqRotatedInteractionHamiltonian}) \cite{Doucot2002,Tovmasyan2018}. Additionally, one can define the global CLS number parity as the sum of the local operators in all unit cells, $\hat{\mathcal{P}}=\sum_k \hat{\mathcal{P}}_k$. Given that the rotated Hamiltonian commutes with the local operator, it is straightforward to see that it also commutes with the global CLS number parity, $[\hat{\mathcal{H}}_n^\prime,\hat{\mathcal{P}}]=0$. The eigenvalues of the global parity are determined by the number of unit cells and the number of particles that can occupy the CLSs. If there are at least as many particles, $N$, as unit cells, $N\geq N_c$, there are $N_c+1$ sectors with eigenvalues 
$\mathcal{P}=-N_c,-N_c+2,...,N_c-2,N_c$. For $N<N_c$, the number of sectors reduces to $N+1$ as the lowest eigenvalues become unavailable. We note that the $\mathbb{Z}_2$ local gauge symmetry makes the sub-sectors in this model similar to the superselection sectors present in lattice gauge theories, where the shattering of the Hilbert space naturally stems from the gauge field and leads to non-ergodicity \cite{Metavitsiadis2017,Smith2018,Russomanno2020a}.

Spinless fermions in diamond lattices with nearest interactions present a locally fragmented Hilbert space where the number of particles occupying a CLS is conserved, which corresponds to a $U(1)$ local gauge symmetry \cite{Danieli2022}. The authors note there that for bosons or spinful fermions, the two-particle tunnelling implies that the number of particles in a CLS is no longer conserved. Here, we show that for bosons with on-site interactions a new conserved quantity emerges, the CLS number parity, which preserves the fragmentation of the Hilbert space. 

In Fig.~\ref{FigGraphs}, we represent three examples of the adjacency graph of the rotated Hamiltonian. Unless otherwise specified, we consider open boundary conditions, and for all simulations we fix $U/\mathcal{J}=1$ and consider an integer number of unit cells. Henceforward, the eigenvalues $\mathcal{P}_k$ are denoted as $\pm$. The width of the lines indicates the strength of the couplings between basis states and, for Figs.~\ref{FigGraphs}(a) and (b), the color of the nodes indicates the diagonal terms of the rotated Hamiltonian, $\varepsilon=\langle f|\hat{\mathcal{H}}_n^\prime|f\rangle$, where $|f\rangle$ is a basis state. Fig.~\ref{FigGraphs}(a) represents the diamond chain, $\hat{\mathcal{H}}_1^\prime$, a known square-root topological insulator \cite{Zurita2021,Marques2021}, with $N=2$ particles in $N_c=2$ unit cells.  Each basis state is represented by a node and labeled using the notation $|N_{C,1}\,N_{S,1}\,N_{A,1}\,N_{C,2}\,N_{S,2}\,N_{A,2}\rangle$, where $N_{j,k}$ is the number of particles in state $|j_k\rangle$ ($j=C,S,A$) in the unit cell $k$. We obtain several uncoupled clusters of basis states with distinct local eigenvalues $\bm{\mathcal{P}}$, \textit{i.e.}, the Hilbert space is fragmented. Each sector with global eigenvalue $\mathcal{P}$ is composed of one or more uncoupled sub-sectors with eigenvalues $\bm{\mathcal{P}}$. There is a one-dimensional (or frozen) sub-sector with a single basis state with the two particles occupying the two CLSs, $\mathcal{P}=-2$, and which is not coupled to any other basis state. There are two sub-sectors sharing the same global CLS parity value, $\mathcal{P}=0$, where only one particle is in a CLS, while the other particle occupies the dispersive chain. The two sub-sectors arise due to the two CLSs that the particle can occupy, which leads to different orderings in the elements of the vector $\bm{\mathcal{P}}$. Finally, most of the basis states of the largest sub-sector have the two particles in the dispersive chain and zero in a CLS. However, due to the two-particle tunnelling, there are two special basis states with two particles occupying the same CLS, $|002000\rangle$ and $|000002\rangle$, which yield the same eigenvalue for the local and global CLS number parity, $\bm{\mathcal{P}}=(+,+)$ and $\mathcal{P}=2$. 

In Fig.~\ref{FigGraphs}(b), we present the same system, $\hat{\mathcal{H}}_1^\prime$, for a larger lattice: $N_c=3$ unit cells with $N=3$ particles. The number of sub-sectors proliferates due to the presence of an additional CLS in the lattice. More precisely, for $N\geq N_c$, the number of sub-sectors is given by $2^{N_c}$, while for $N<N_c$, the number is $\sum_{k=0}^N\binom{N_c}{k}$. The sub-sectors $\mathcal{P}=-1$, like the sub-sectors $\mathcal{P}=0$ in Fig. \ref{FigGraphs}(a), have only one particle in the dispersive chain, while all the other particles occupy distinct CLSs. These particles can not access the two-particle tunnelling in Eq. (\ref{EqRotatedInteractionHamiltonian}) and thus, are trapped in the CLSs. Therefore, these sub-sectors are effectively single-particle systems with a non-uniform on-site potential distribution. For the larger sub-sectors, there are at least two particles in the dispersive chain, making these  sub-sectors sensitive to interactions. Note that the different sub-sectors with the same eigenvalue $\mathcal{P}$ for the global CLS number parity are not degenerate due to the different positioning of the diagonal terms. This will prove to be an important factor in distinguishing between the different domes of the distribution of entanglement entropies, as we discuss below in Sec.~\ref{SecModelComparison}. 

As an example of a sub-sector with a large dimension, we represent the largest sub-sector of $\hat{\mathcal{H}}_1^\prime$, with $N=4$ particles in $N_c=4$ unit cells in Fig.~\ref{FigGraphs}(c). The color of each basis state represents the total number of particles that are in a CLS, $N_{CLS}|f\rangle=\sum_k \hat{n}_{a,k}|f\rangle$. Most of the basis states have zero particles in a CLS. However, there are also some basis states with four-particles in a CLS (either four-particles in the same CLS or two pairs of particles in different CLSs), and many-more with two particles in the same CLS. This embedding of special basis states has some consequences on the distribution of entanglement entropies of the system, which will be discussed in Section \ref{SecModelComparison}.

It is important to note that the Hilbert space fractures into a series of uncoupled sub-sectors only on the rotated or entangled basis. Meanwhile, the Hilbert space in the original or product-state basis exhibits a connected adjacency graph. Thus, the results of this section show how this system exhibits quantum Hilbert space fragmentation, a distinction recently proposed in \cite{Moudgalya2021b}. In contrast, the Hilbert space of classically fragmented systems is shattered in the product-state basis. While the fracture is only revealed on the rotated basis, it still has some dramatic consequences on the thermalization properties of this family of models, which we explore in Section \ref{SecNumericalResults}. 

Another recently proposed classification of Hilbert space fragmentation distinguishes between strongly and weakly fragmented systems  in the context of dipole conserving models \cite{Sala2020,Khemani2020,Morningstar2020}. The ratio between the dimension of the largest sector $\mathscr{D}_{max}$ and the dimension of the full Hilbert space $\mathscr{D}$ either tends to one in the thermodynamic limit, signaling weak fragmentation, or tends to zero, signaling strong fragmentation. Typical initial states of a weakly fragmented system belong to the largest sector, and thus, completely thermalize, while only a vanishing subset of initial states are non-thermal. For strongly fragmented systems, most initial states only have access to a small subset of the Hilbert space, which precludes full thermalization. Thus, these two types of fragmentation are associated with a weak or a strong violation of the ETH, respectively. For our model, the dimension of the largest sector is given by
\begin{equation}
	\mathscr{D}_{max}\hspace{-0.2mm}=\hspace{-0.3mm}\sum_{\varrho=0}^{\lfloor N/2\rfloor}\hspace{-1mm}\binom{(n+1)N_c+(N-2\varrho)-1}{N-2\varrho}\hspace{-0.5mm}
	\binom{N_c+\varrho-1}{\varrho}\hspace{-0.2mm},
\end{equation} 
where the integer $\varrho$ counts the number of pairs of particles that populate the CLSs. Given the dimension of the full Hilbert space,
\begin{equation}
	\mathscr{D}=\binom{(n+2)N_c+N-1}{N},
\end{equation}
the ratio $\mathscr{D}_{max}/\mathscr{D}$ tends to zero at the thermodynamic limit, indicating strong Hilbert space fragmentation for this family of models. Thus, this result points to a strong violation of the ETH, as we will numerically argue in the next Section.

Finally, this system exhibits local Hilbert space fragmentation, a term recently coined in \cite{Buca2022}, as the fragmentation stems from a local conservation law, namely, the local CLS number parity, $[\hat{\mathcal{H}}_n^\prime,\hat{\mathcal{P}}_k]=0$.

\begin{figure}[t]
	\includegraphics[width=0.95\columnwidth]{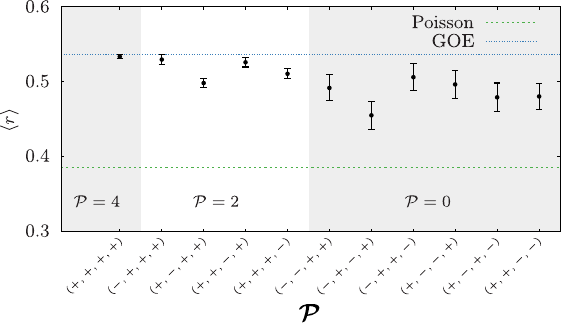}
	\caption{Mean level spacing ratio for the sub-sectors with $\mathcal{P}=4,2,0$ of $\hat{\mathcal{H}}_4^\prime$, with $N=4$ particles in $N_c=4$ unit cells. The blue dotted line indicates the value corresponding to the Gaussian orthogonal ensemble, $\langle r\rangle_{GOE}=0.536$, and the dashed green line, the value for a Poisson distribution, $\langle r\rangle_P=0.386$. The error bars are standard errors of the mean. The sub-sectors $\mathcal{P}=-2$ and $\mathcal{P}=-4$, which are not included, correspond to the integrable effective single-particle sub-sectors and the integrable frozen states, respectively.}\label{FigMeanLevelSpacing}
\end{figure}

\section{Exact diagonalization results}\label{SecNumericalResults}

\subsection{Level statistics}

In order to characterize the properties of the different sub-sectors of the Hilbert space, we analyze their level statistics using exact diagonalization. For each sub-sector, we consider the ordered eigenvalues $E_n$, and the nearest-neighbor gaps $s_n=E_{n+1}-E_{n}$. From those, one can define the level spacing ratios for each pair of gaps \cite{Oganesyan2007},
\begin{equation}
	r_n=\frac{\min \left(s_{n}, s_{n+1}\right)}{\max \left(s_{n}, s_{n+1}\right)},
\end{equation}
and the corresponding average $\langle r\rangle$. Non-integrable systems with time-reversal symmetry are expected to approximate the probability distribution $P(r)$ of the Gaussian orthogonal ensemble, with an average value $\langle r\rangle_{GOE}=0.536$ \cite{Atas2013}. For integrable systems, a Poisson distribution is expected, with a characteristic value  $\langle r\rangle_{P}=0.386$. In Fig.~\ref{FigMeanLevelSpacing}, we represent the average spacing ratio for $\hat{\mathcal{H}}_4^\prime$ with $N=4$ particles in $N_c=4$ unit cells for the sub-sectors $\mathcal{P}=4,2,0$. We observe how most sub-sectors are within a few error bars of $\langle r\rangle_{GOE}$. The value of $\langle r\rangle$ increases with the global CLS number parity, $\mathcal{P}$, as less particles are trapped in a CLS. Additionally, the lowest values of $\langle r\rangle$ correspond to the sub-sectors with the smallest dimension (\textit{i.e.} smaller $\mathcal{P}$), for which the $P(r)$ distribution is not so well-defined. Besides the sub-sectors shown in Fig. \ref{FigMeanLevelSpacing}, the system also presents the integrable sub-sectors with $\mathcal{P}=-2$, the effectively single-particle sub-sectors, and $\mathcal{P}=-4$, the frozen, one-dimensional sub-sector.

\begin{figure}[t]
	\includegraphics[width=1\columnwidth]{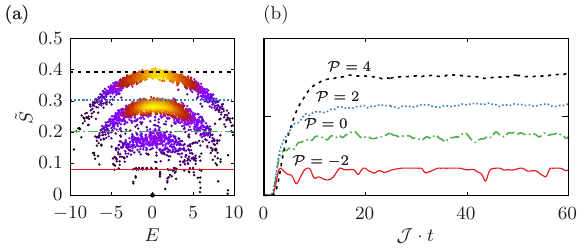}
	\caption{Distribution of entanglement entropies and entanglement evolution of a trial state for $\hat{\mathcal{H}}_2^\prime$, with $N=4$ particles in $N_c=4$ unit cells. (a) Half-chain bipartite von Neumann entanglement entropy of each eigenstate as a function of the energy. The horizontal lines are the sector-restricted Page values for each sector and the color of the dots indicates the density of data points, increasing with warming colors. (b) Entanglement evolution of a rotated basis state of each sector with eigenvalue $\mathcal{P}$. In both figures, the entropy is normalized to the number of sites in the subsystem, $\tilde{S}=S/\mathcal{N}_L$.}\label{FigPageAndEvolution}
\end{figure}

\subsection{Entanglement entropy and evolution}
In this Section, we calculate the bipartite von Neumann entanglement entropy, $S$, for each eigenstate of the full Hilbert space by partitioning the lattice into two subsystems: left, $L$; and right, $R$. The entanglement entropy is then $S=-\operatorname{tr}(\rho_L \ln \rho_{L})$, where $\rho_L$ is the reduced density matrix of the left subsystem. We consider the half-chain entanglement entropy by partitioning the lattice in the middle, with the same number of sites in each subsystem and such that the cut never falls between the $U$ and $D$ sites of a single diamond.   

Fig.~\ref{FigPageAndEvolution}(a) represents the entanglement entropy of all eigenstates of the system for $\hat{\mathcal{H}}_2^\prime$ with $N=4$ particles in $N_c=4$ unit cells. We give the results in terms of the normalized entanglement entropy, $\tilde{S}=S/\mathcal{N}_L$ where $\mathcal{N}_L$ is the number of sites subsystem $L$. The entanglement entropy is not a continuous function of the energy density but presents a nested-dome structure. Similar structures have been found in the distribution of entanglement entropies of systems with \cite{McClarty2020,Lee2020,Chertkov2021} or without frustration \cite{Richter2022}. In Fig.~\ref{FigPageAndEvolution}(a), each dome corresponds to a sector with a given value for the global CLS number parity, $\mathcal{P}$. 

\begin{itemize}
	\item The upper dome corresponds to the largest sub-sector, with $\mathcal{P}=4$, where most basis states have all particles in the dispersive chain and none is trapped in a CLS.
	\item The second dome from above corresponds to the sub-sectors with $\mathcal{P}=2$, where most basis states have one particle in a CLS and the other three are in the dispersive chain. As the subsystem partition does not fall between the sites $U$ and $D$ of any diamond, the contribution to the entanglement of the particle occupying a CLS is exactly zero. Thus, the eigenstates belonging to the sector $\mathcal{P}=2$ have an upper bound for the entanglement entropy given by the maximum number of particles in the dispersive chain of the corresponding basis states.
	\item The third dome corresponds to $\mathcal{P}=0$, where most basis states have two particles in a CLS and two in the dispersive chain. Consequently, those sub-sectors have an even lower bound for the entanglement entropy. 
	\item The sub-sectors with $\mathcal{P}=-2$ have only one particle in the dispersive chain, making them effectively single-particle systems. As a result, their distribution of entanglement entropies does not form a dome structure. Most eigenstates accumulate at a constant value, which one would expect for a linear chain, while some fall below as a consequence of the interaction-induced on-site potentials, \textit{e.g.}, a particle occupying the $|A_k\rangle$ CLS can be translated into an effective on-site potential of strength $U$ acting on a second particle located at $|S_k\rangle$ of the dispersive chain, due to the first term of $\mathcal{H}^{'int}_{n,diam.}$ in Eq.~(\ref{EqRotatedInteractionHamiltonian}). These potentials act as impurities that either attract or repel the wavefuntions, and they induce an asymmetry between the $L$ and $R$ subsystems that lowers the half-chain  entanglement entropy. 
	\item Finally, there is a single state with exactly zero entanglement entropy and zero energy that corresponds to the sub-sector with $\mathcal{P}=-4$, for which all the particles are trapped in a CLS. 
\end{itemize}
 
The interaction-induced on-site potentials are the origin of the many-body localization transition observed in \cite{Daumann2020,Danieli2022} for the diamond chain with nearest-neighbor interactions and spinless fermions. For spinless fermions, the two-particle tunnelling is not present, thus completely decoupling the CLSs from the dispersive chain, and the random on-site potentials cause a transition to a many-body localized phase when the interaction, \textit{i.e.}, the effective disorder, is increased. 

Additionally, we plot in Fig.~\ref{FigPageAndEvolution}(a) the sector-restricted Page value (horizontal lines), for each of the sectors with a given global CLS number parity $\mathcal{P}$. The Page value is the average entanglement entropy of a random vector, for which Don N. Page derived an analytical expression for a generic bipartite quantum system \cite{Page1993}. We find the Page value using normalized random vectors $|\psi\rangle$ of the form
\begin{equation}
	\langle f \mid \psi\rangle=\frac{\alpha_{f, \psi}+i \beta_{f, \psi}}{\mathcal{N}_{\psi}},
\end{equation}
where the basis states $|f\rangle$ belong to a particular sub-sector $\bm{\mathcal{P}}$, $\alpha_{f, \psi}$ and $\beta_{f, \psi}$ are taken from a normal distribution with zero mean, and $\mathcal{N}_{\psi}$ is the normalization constant. The entanglement entropy of each random state is computed by projecting $|\psi\rangle$ onto the full Hilbert space. Then, we compute the average of the entanglement entropy for one thousand random vectors belonging to a particular sub-sector, such that the sector-restricted Page value is given by the average value of the corresponding sub-sectors. Each sector-restricted Page value coincides with the top of each dome [see Fig.~\ref{FigPageAndEvolution}(a)]. In Fig.~\ref{FigPageAndEvolution}(b), we take a basis state belonging to each sector and let it evolve through time (in dimensionless units, $\mathcal{J}\cdot t$). The evolved wavefunctions are computed numerically using the time-evolution unitary operator defined through the Hamiltonian, Eqs.~(\ref{EqSingleRotatedParticleHamiltonian}) and (\ref{EqRotatedInteractionHamiltonian}). In particular, we take the following states as an example 
\begin{equation}
	\begin{aligned}
		\mathcal{P}&=4 &\qquad  &|0000000000000004\rangle,\\
		\mathcal{P}&=2 &\qquad  &|3000000000010000\rangle, \\
		\mathcal{P}&=0 &\qquad  &|2001000100000000\rangle,  \\
		\mathcal{P}&=-2 &\qquad &|1001000100010000\rangle. \\
	\end{aligned}
\end{equation}
We observe how the entanglement entropy at which the evolved state saturates is bounded by the corresponding sector-restricted Page value indicated in Fig \ref{FigPageAndEvolution}(a). This sector-restricted weak thermalization induced by the fragmentation of the Hilbert space constitutes a violation of the ETH.

\subsection{Entanglement scaling}
\begin{figure}[t]
	\includegraphics[width=1\columnwidth]{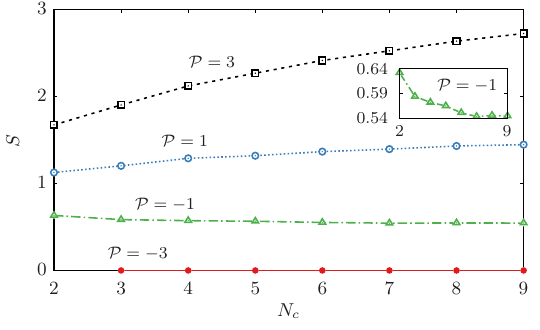}
	\caption{Average entanglement entropy for each sector $\mathcal{P}$ as a function of the number of unit cells $N_c$ for $\hat{\mathcal{H}}_2^\prime$ with $N=3$ particles. The inset shows sector $\mathcal{P}=-1$. The lines are represented as a guide to the eye.}\label{FigEntanglementGrowth}
\end{figure}
\begin{figure*}[t]
	\includegraphics[width=2\columnwidth]{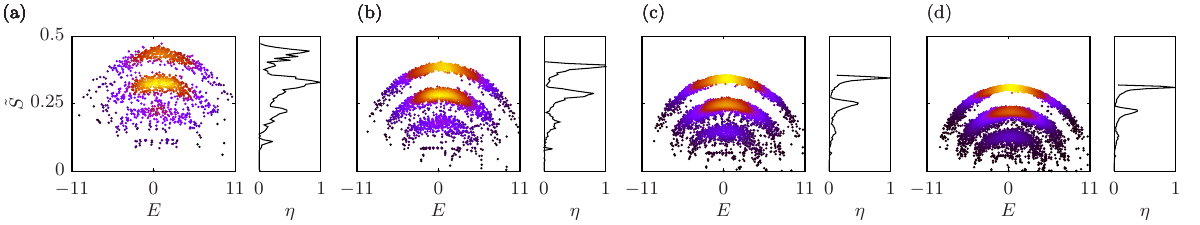}
	\vspace{0mm}
	\includegraphics[width=2\columnwidth]{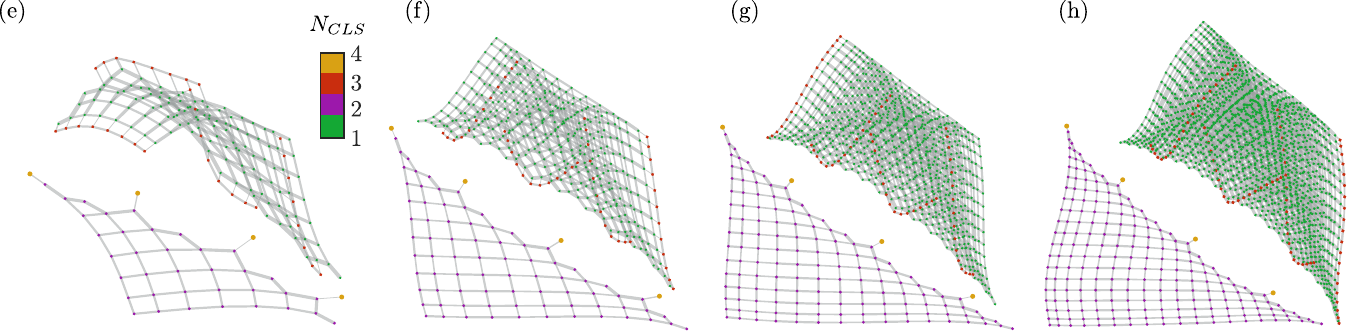}
	\caption{Distribution of entanglement entropies and adjacency graphs for $N=4$ particles in $N_c=4$ unit cells for the following models: (a),(e) $\hat{\mathcal{H}}_1^\prime$; (b),(f) $\hat{\mathcal{H}}_2^\prime$; (c),(g) $\hat{\mathcal{H}}_3^\prime$; (d),(h) $\hat{\mathcal{H}}_4^\prime$. (a)-(d) Left plots: normalized entanglement entropy $\tilde{S}$ as a function of the energy, where color represents the density of data points. (a)-(d) Right plots: normalized density of data points as a function of $\tilde{S}$ for the middle region of the spectrum, $-|E_{0}|\cdot0.2<E<|E_{0}|\cdot0.2$. Plots (e)-(h): second and third largest sub-sectors in the adjacency graphs of the rotated Hamiltonians with the color of the nodes indicating the total number of particles occupying a CLS, $N_{CLS}$.}\label{FigModelComparison}
\end{figure*}
In order to further characterize the properties of the different sectors of the Hamiltonian, we compute the scaling of the entanglement entropy $S$ with system size for each of the sectors $\mathcal{P}$. In Fig.~\ref{FigEntanglementGrowth}, we plot the average entanglement entropy for the eigenstates of each sector as one increases the number of unit cells of the Hamiltonian $\hat{\mathcal{H}}_2^\prime$ with $N=3$ particles. The sectors $\mathcal{P}=3$ and $\mathcal{P}=1$ exhibit logarithmic entanglement growth, thus demonstrating subthermal behavior within each sector \cite{Moudgalya2018,Papic2021}. However, the growth rate of both sectors is different, as most basis states in sector $\mathcal{P}=3$ contain three particles in the dispersive chain while none are trapped in a CLS. In contrast, most basis states in sector $\mathcal{P}=1$ only have two particles in a dispersive state while one is trapped in a CLS. Both sectors present a logarithmic growth of the form $S=\sigma \ln(N_c)+\upsilon$ with $\{\sigma=0.708\pm0.016,\upsilon=1.148\pm0.027\}$ for $\mathcal{P}=3$ and $\{\sigma=0.217\pm0.006,\upsilon=0.973\pm0.010\}$ for $\mathcal{P}=1$. 

The sector $\mathcal{P}=-1$ corresponds to the effectively single-particle sub-sectors, for which one observes a surprising slight decrease in the entanglement entropy as the size of the system increases (see inset in Fig.~\ref{FigEntanglementGrowth}). This is due to the on-site potential terms that arise in the dispersive chain reflecting the presence of one particle in $|A_k\rangle$ and one in $|S_k\rangle$. Any left-right subsystem asymmetries in the location of the two nodes of the adjacency graph with an on-site potential will lower the entanglement entropy. For $N_c=2$, there is a single sub-sector where the two basis states that have an on-site potential fall in opposite subsystems $L$ and $R$. As the size of the system increases, more CLSs are available and thus there are more sub-sectors where there is some asymmetry in the location of the on-site potential (\textit{e.g.} the two potentials may fall in the same subsystem $L$ or $R$). Thus, the average entanglement entropy of the sector $\mathcal{P}=-1$ slightly decreases with system size. The decrease is more pronounced for small numbers of unit cells, and it seems to tend to an asymptotic value. This constitutes an anti-volume correction that should also play a role in sectors $\mathcal{P}=1,3$, though it is not noticeable there as the logarithm term dominates. 

The sector with $\mathcal{P}=-3$ includes the one-dimensional sub-sectors where all the particles are trapped in a CLS. This sector follows an area law scaling, which in one dimension corresponds to a constant value. As a particle in a CLS does not contribute to the entanglement entropy, the average entanglement entropy for these sub-sectors is zero for any system size. If one diagonalizes the full Hilbert space in the original basis to compute the entanglement entropy for this sector, one obtains a series of degenerate states that correspond to the different CLSs that the three particles can occupy. Then, the entanglement entropy obtained through this method is higher than the one shown in Fig. \ref{FigEntanglementGrowth}, as it corresponds to an arbitrary numerical superposition of those states. Consequently, one should compute the entanglement entropy of this sector in the rotated basis. Note that this sector does not exist for $N_c=2$ unit cells: as the number of particles is $N=3$, one particle will always occupy the linear chain. 

These results demonstrate that the system exhibits weak thermalization with respect to the full Hilbert space through its fragmentation, while also exhibiting subthermal behavior within each non-integrable sector.

\subsection{Model comparison and boundary conditions}\label{SecModelComparison}
In this subsection, we analyze the effect that the number of central sites of the lattice has in the distribution of entanglement entropies by comparing the different models of the family of diamond necklaces. Fig. \ref{FigModelComparison} shows the entanglement spectra and the adjacency graphs of different models for $N=4$ particles in $N_c=4$ unit cells. The represented models are: (a),(e) $\hat{\mathcal{H}}_1^\prime$; (b),(f) $\hat{\mathcal{H}}_2^\prime$; (c),(g) $\hat{\mathcal{H}}_3^\prime$; (d),(h) $\hat{\mathcal{H}}_4^\prime$. The left subplots in the upper row show the normalized entanglement entropy $\tilde{S}$ for each eigenstate as a function of the energy. The color indicates the density of data points. The right subplots in the upper row represent the density of data points $\eta$ as a function of $\tilde{S}$ for the eigenstates around $E=0$. To obtain a clear picture, we take the eigenstates whose energy fulfills  $-|E_{0}|\cdot0.2<E<|E_{0}|\cdot0.2$, where $E_0$ is the ground-state energy, and normalize the density $\eta$ to $1$. The lower row of plots show the second and third largest sub-sectors in the adjacency graph of the rotated Hamiltonian. The color of the nodes indicates the total number of particles that occupy a CLS, as given by $N_{CLS}|f\rangle=\sum_k \hat{n}_{a,k}|f\rangle$. 

We see how increasing the number of central sites in the lattice, going from $\hat{\mathcal{H}}_1^\prime$ to $\hat{\mathcal{H}}_4^\prime$, increases 
the visibility of the different domes. This can be understood in terms of the adjacency graphs of the different models. In Figs.~\ref{FigModelComparison}(e)-(h), most of the basis states of the lower row of sub-sectors have two particles in different CLSs, in purple, although there are some special basis states, in yellow, where an additional pair of particles also occupies a CLS. A similar pattern occurs in the sub-sectors of the upper row, for which most basis states have one-particle in a CLS, in green, while some have three particles occupying CLSs, in red. These special basis states appear due to the two-particle tunnelling term in the rotated Hamiltonian of Eq.~(\ref{EqRotatedInteractionHamiltonian}), and thus are present in all sectors except for the integrable ones. The eigenstates that have some weight on those basis states will have a lower entanglement entropy than those that do not, and they might fall below the dome of the sub-sector, thus obscuring the visibility of the nested-dome pattern. When one increases the number of central sites in the lattice, these special basis states become more sparse compared to the main basis states, which have a lower number of particles in a CLS [see Figs.~\ref{FigModelComparison}(e)-(h)]. Therefore, the visibility of the nested-dome structure in the distribution of entanglement entropies can be enhanced by increasing the sparsity of the CLSs. This, in turn, increases the sparsity of the special basis states with a higher number of particles in a CLS due to the two-particle tunnelling. 

Let us consider what would occur for different numbers of particles. For each particle added with respect to a fixed number of unit cells, an extra dome appears on top and one dome (or sector) is removed from below. For example, for $N=5$ and $N_c=4$ unit cells, the frozen sub-sector is unavailable. However, the number of domes for $N\geq N_c$ is conserved, as it corresponds to the number of sectors. As one increases the number of particles, there is a global shift to the right in the distribution of entanglement entropies, which corresponds to an increased energy of the eigenstates due to the repulsive interaction. For each particle removed, keeping the number of unit cells fixed, the upper dome disappears, as there are less particles populating the dispersive chain. Additionally, the frozen sub-sectors multiply, due to the different CLSs that the particles can occupy, and become degenerate.

Up to now, we have assumed open boundary conditions, however, this analysis also holds for periodic boundary conditions. The visibility of the domes when one introduces periodic boundary conditions is notably worse than for open boundary conditions. This is due to the fact that periodic boundary conditions make the system translation invariant, which introduces degeneracies in the spectrum between sub-sectors belonging to the same sector. As a result, one numerically finds arbitrary superpositions of the degenerate eigenstates which have arbitrary entanglement entropies. The cause of the deteriorated visibility can be corroborated by introducing vertical couplings between the $U$ and $D$ sites of each diamond and making their strength different for each unit cell. In that case, although the system still has periodic boundary conditions, it is no longer translation invariant, and the visibility of the domes is restored.

\section{Conclusions}\label{SecConclusions}
We have studied Bose-Hubbard models in a family of diamond necklace lattices with $n$ central sites. Such models possess a single-particle spectrum with a flat band, which is composed of compact localized states (CLSs) located in each diamond. Due to the presence of these CLSs, when adding more bosons with on-site interactions, the Hilbert space becomes locally fragmented. We have demonstrated how this fragmentation is revealed in the adjacency graph of the Hamiltonian when applying an appropriate basis rotation that decouples the CLSs at the single-particle level, making it an instance of quantum local Hilbert space fragmentation. Also, by analyzing the dimension of the largest sector, we have shown that the system exhibits strong fragmentation, which leads to a strong violation of the Eigenstate Thermalization Hypothesis. We have found a conserved quantity that uniquely identifies each sub-sector of the Hamiltonian, the local CLS number parity. The sub-sectors present a wide range of dimensions, including one-dimensional sub-sectors, and also entanglement entropy scalings ranging from area-law to logarithmic growth, while also including one sector with an anti-volume correction. As a result of the fragmentation, the distribution of entanglement entropies presents a nested-dome structure, that stems from the number of particles that are trapped in a CLS. We have found weak thermalization through sub-sector-restricted entanglement evolution and subthermal entanglement growth within each non-integrable sector. Additionally, we have shown how the visibility of the nested-dome structure can be enhanced by increasing the sparsity of the CLSs, and how the results hold both for open and periodic boundary conditions. 

These results can be generalized to higher-dimension versions of the diamond necklace while another interesting extension of this work is the study of other flat-band models, as these systems have been realized in a variety of experimental platforms (see \cite{Leykam2018} and references therein). Lattices supporting orthogonal CLSs can be detangled into a dispersive lattice and a series of decoupled CLSs \cite{Flach2014a}, thus already providing the first ingredient for many-body Hilbert space fragmentation and weak fragmentation.

\section{Acknowledgments}
EN, VA, and JM acknowledge support through the grant PID2020-118153GB-I00 funded by MCIN/AEI/ 10.13039/501100011033, the Catalan Government (Contract No. SGR2017-1646), and the European Union Regional Development Fund within the ERDF Operational Program of Catalunya (project QUASICAT/QuantumCat). EN acknowledges financial support from MCIN/AEI/ 10.13039/501100011033 through the grant PRE2018-085815 and from COST through Action CA16221. AMM and RGD acknowledge financial support from the Portuguese Institute for Nanostructures, Nanomodelling and Nanofabrication (i3N) through Projects No. UIDB/50025/2020, No. UIDP/50025/2020, and No. LA/P/0037/2020, and funding from FCT–Portuguese Foundation for Science and Technology through Project No. PTDC/FISMAC/29291/2017. AMM acknowledges financial support from the FCT through the work Contract No. CDL-CTTRI147-ARH/2018 and from i3N through the work Contract No. CDL-CTTRI-46-SGRH/2022.

\end{document}